\documentclass{PoS}

\usepackage{amsmath}
\usepackage{graphicx}
\usepackage{amssymb}
\usepackage{subfigure}
\usepackage{cancel}
\usepackage{relsize}
\usepackage{bbold}
\usepackage{lineno}
\usepackage{braket}
\usepackage{slashed}
\usepackage{multirow}
\usepackage{placeins}
\usepackage{xspace}

\newcommand{\HEP}{\texttt{HEPfit}\xspace}

\title{\boldmath{$B\to K^*\ell^+\ell^-$} in the Standard Model: Elaborations and Interpretations }

\ShortTitle{$B\to K^*\ell^+\ell^-$ in the Standard Model: a Postmortem}

\author{Marco Ciuchini$^a$, 
	    Marco Fedele$^{b,c}$, 
	    Enrico Franco$^c$, 
	    Satoshi Mishima$^d$, 
	    \speaker{Ayan Paul}$^c$, 
	    Luca Silvestrini$^c$ 
	    and Mauro Valli$^{e}$\\	
\llap{}\\	    
\llap{$^a$}INFN, Sezione di Roma Tre, Via della Vasca Navale 84, I-00146 Roma, Italy\\
\llap{$^b$}Dipartimento di Fisica, Universit\`a di Roma ``La Sapienza'', P.le A. Moro 2, I-00185 Roma, Italy\\
\llap{$^c$}INFN, Sezione di Roma, P.le A. Moro 2, I-00185 Roma, Italy\\
\llap{$^d$}Theory Center, IPNS, KEK, Tsukuba 305-0801, Japan\\
\llap{$^e$}SISSA, via Bonomea 265, I-34136 Trieste, Italy\\

E-mail:
\email{marco.ciuchini@roma3.infn.it},
\email{enrico.franco@roma1.infn.it},
\email{marco.fedele@uniroma1.it},
\email{satoshi.mishima@kek.jp},
\email{ayan.paul@roma1.infn.it},
\email{luca.silvestrini@roma1.infn.it},
\email{mauro.valli@roma1.infn.it}}

\abstract{Disentangling New Physics effects from the Standard Model requires a good understanding of all pieces that stem from the latter, especially the uncertainties that might plague the theoretical estimations within the Standard Model. In the light of recent measurements made in the decay of $B\to K^*\ell^+\ell^-$, and accompanying possibilities of New Physics effects, we re-examine the hadronic uncertainties that come about in this exclusive $b \to s$ transition. We show that it is not trivial to distinguish New Physics effects from these hadronic uncertainties and we attempt to quantify the  latter in its magnitude and kinematic shape from the recent LHCb measurements of the angular observables  in this decay mode. We also update our fit with the more recent calculations of the form factors combined with the ones computed with Lattice QCD.
}

\FullConference{38th International Conference on High Energy Physics\\
		3-10 August 2016\\
		Chicago, USA}

\begin{document}

\section{Introduction}
\label{sec:intro}

Much has been discussed and debated about the implications of the measurements made by LHCb~\cite{Aaij:2015oid} and Belle~\cite{Abdesselam:2016llu} experiments of the angular distribution of decay $B^0\to K^{*0}\mu^+\mu^-$. The bone of contention is the stance on whether New Physics (NP) can be disentangled from Standard Model (SM) contributions in these angular observables or not. This boils down to a choice of the prescription for the coherent treatment of uncertainties stemming from the long distance QCD contributions which lie beyond the perturbative and the factorizable regime. We made an attempt~\cite{Ciuchini:2015qxb} to extract this hadronic contribution from data assuming that the data was pointing towards a SM description. We performed this Bayesian analysis with \href{http://hepfit.roma1.infn.it/}{\HEP} and compared what we extracted as the hadronic contribution with what is known as the leading effects~\cite{Khodjamirian:2010vf}. Within reasonable estimates of what can possibly be the nature and size of the hadronic contribution as one gets kinematically close to the $c\bar{c}$ threshold, we concluded that one cannot disentangle NP from SM with significant degree of confidence and, indeed, the latter interpretation is favoured if one takes into consideration the shape of the hadronic contribution. This result is claimed to be at variance with several similar analyses performed assuming that the leading single soft gluon contribution advocated in~\cite{Khodjamirian:2010vf} is sufficient for the entire kinematic range in the large recoil (low $q^2$) regime~\cite{Altmannshofer:2014rta,Descotes-Genon:2015uva}, albeit, with accommodations made for the phase of this contribution but with no, possibly significant, enhancement to its size close to the $c\bar{c}$ threshold. 

In this proceeding, we focus on the procedure of estimation and the interpretation of the size of the hadronic contribution that we extract from the data as compared to the other analogous contributions that add up to the total amplitude. We also discuss the functional form of this hadronic contribution and explain how it can be different from what NP at higher scales contributing through shifts in the short distance (SD) Wilson coefficients would generate. We take this opportunity to reevaluate the observables with updated form factors~\cite{Straub:2015ica} which can now be coherently combined with Lattice calculations~\cite{Horgan:2015vla} of the same at low recoil.

\section{On the size of the hadronic contribution}
\label{sec:hadronic}
\begin{figure}[!h]
\centering
\subfigure[$2C_2\left|\tilde{g}_{i}^{fit}\right|$ extracted from LHCb data when constraining $\tilde{g}_{i}$ using estimates from~\cite{Khodjamirian:2010vf} below $q^2=1$ GeV$^2$.]{\includegraphics[width=0.9\textwidth]{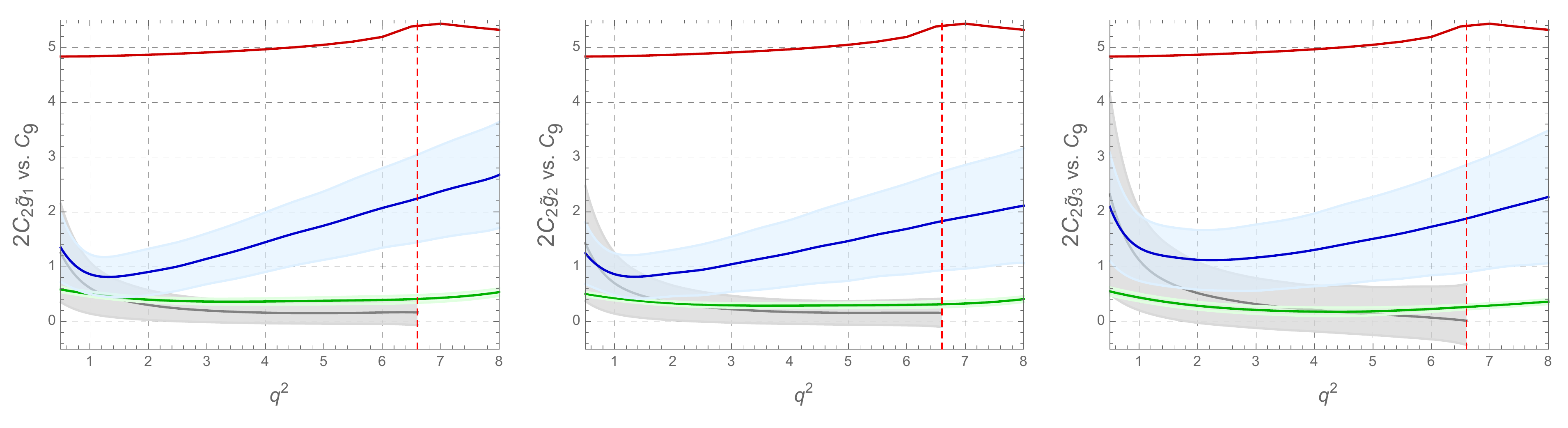}}\\
\subfigure[$2C_2\left|\tilde{g}_{i}^{fit}\right|$ extracted from LHCb data without using any theoretical estimation of $\tilde{g}_{i}$.]{\includegraphics[width=0.9\textwidth]{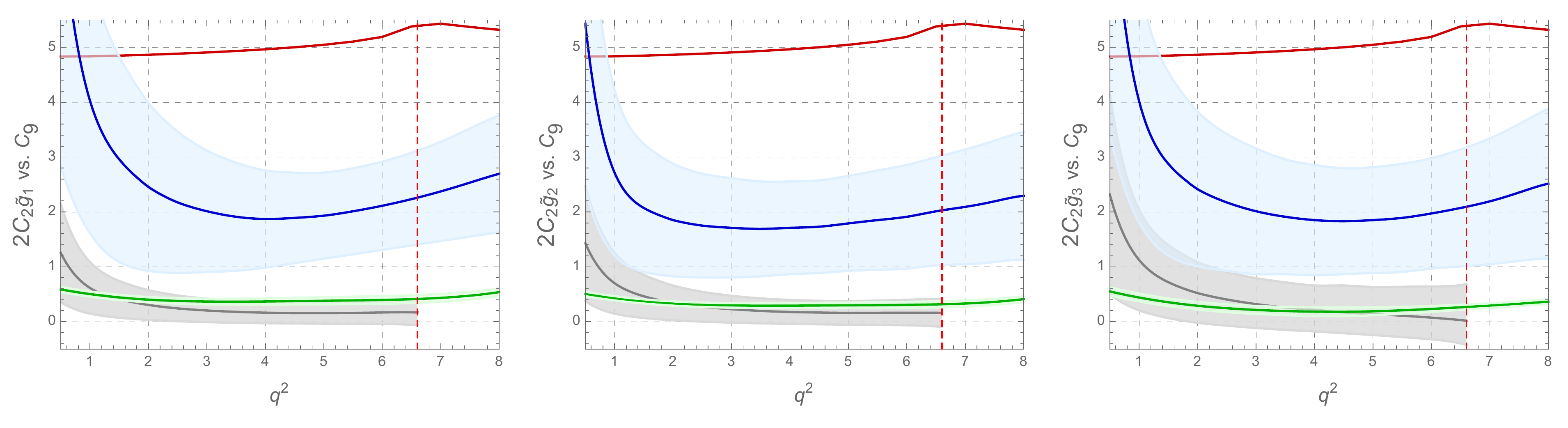}}\\
\subfigure[Same as (a) but with $h_\lambda^{(2)}=0$, hence removing the $q^4$ dependence of the non-factorizable contribution.]{\includegraphics[width=0.9\textwidth]{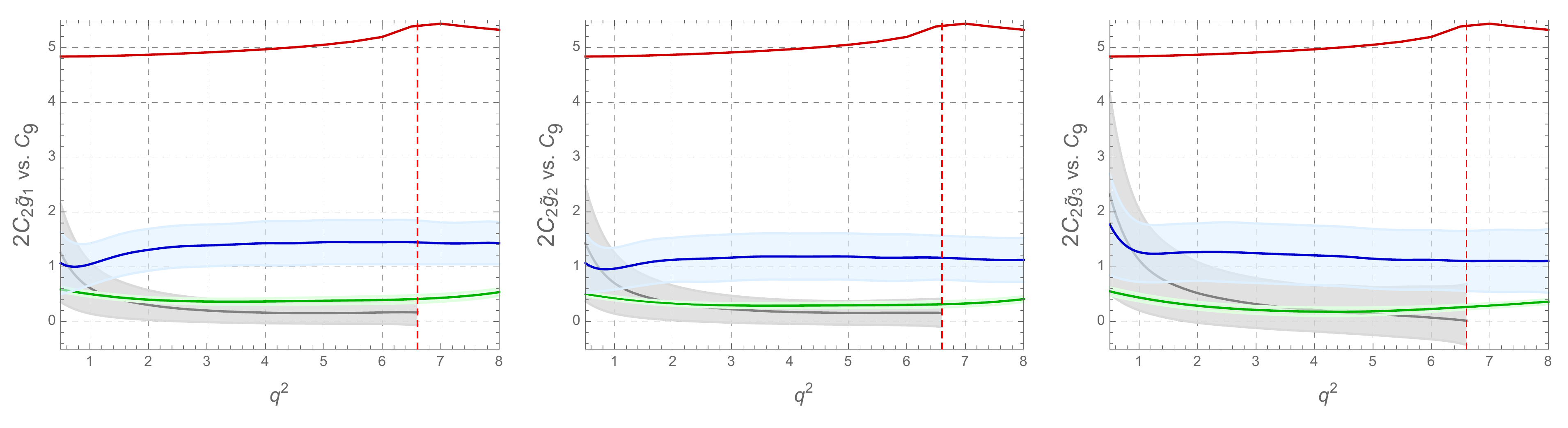}}\\
\subfigure{\includegraphics[width=0.6\textwidth]{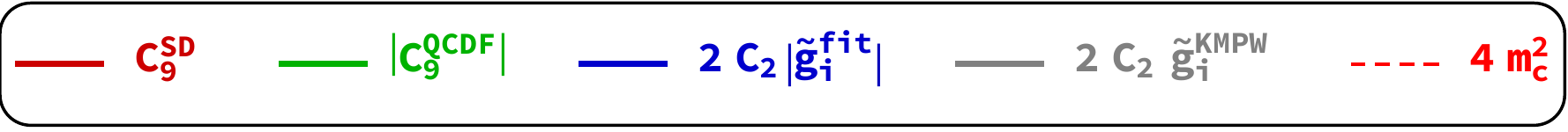}}
\caption{Comparison of $2C_2\vert\tilde{g}_{i}^{fit}\vert$ (blue band) extracted  from our fit to the LHCb data with $C_{9}^{SD}=C_9(\mu_b)+Y(q^2)$ (red line) with  $\mu_b=4.8$ GeV and the contribution to $\vert C_9^{QCDF}\vert $ (green band). The gray band is $2C_2\tilde{g}_{i}^{KMPW}$ estimated from~\cite{Khodjamirian:2010vf}. The thick lines are the central values and the lighter bands the RMS of the distribution.}
\label{fig:comp}
\end{figure}

Considering the complexities of estimating the hadronic uncertainties from first principles or by some approximate methods, one can try to extract these from data and compare their size to other factorizable and SD contributions to estimate the legitimacy of their magnitude. In our work~\cite{Ciuchini:2015qxb} we have elucidated why we think the non-factorizable contributions presented in~\cite{Khodjamirian:2010vf} might be underestimated as one gets close to the $c\bar{c}$ threshold. In fig.~\ref{fig:comp} we present a comparison between the hadronic uncertainties extracted from the LHCb data~\cite{Aaij:2015oid} and the other SD and factorizable contributions. The red lines show the SD contribution $C_{9}^{SD}=C_9(\mu_b)+Y(q^2)$ with $\mu_b=4.8$ GeV, where $Y(q^2)$ denotes the perturbative charm loop contribution. The green bands are the QCDF corrections, $\vert C_{9}^{QCDF}\vert$ with the pole at vanishing $q^2$ subtracted out. The blue bands show $2C_2\vert\tilde{g}_{i}^{fit}\vert$, with $i=1,2,3$, as extracted from LHCb data which can be compared to the gray bands showing $2C_2\tilde{g}_{i}^{KMPW}$ extracted from~\cite{Khodjamirian:2010vf}. It is clear that the non-factorizable contributions are significantly smaller than the SD contribution, even as one gets close to the $c\bar{c}$ threshold, while being larger than the QCDF contribution. While in~\cite{Khodjamirian:2010vf} $\tilde{g}_i$ is real, we allow for it to have an imaginary part and hence the plots in fig.~\ref{fig:comp} show $\vert\tilde{g}^{fit}_i\vert$.

\section{On the functional form of the hadronic contribution}
\label{sec:intro}

Not only is the size of the hadronic contribution,  $h_\lambda(q^2)$, important but so its its functional form which we expand to $\mathcal{O}(q^4)$:
\begin{equation}
 h_\lambda(q^2) = \frac{\epsilon^*_\mu(\lambda)}{m_B^2} \int d^4x e^{iqx} \langle \bar K^* \vert T\{j^{\mu}_\mathrm{em} (x) 
 \mathcal{H}_\mathrm{eff}^\mathrm{had} (0)\} \vert \bar B \rangle
 = h_\lambda^{(0)} + \frac{q^2}{1\,\mathrm{GeV}^2}
         h_\lambda^{(1)} + \frac{q^4}{1\, \mathrm{GeV}^4} h_\lambda^{(2)} + \mathcal{O}(q^6).
 \label{eq:hlambda}
\end{equation}
The relation between $ h_\lambda(q^2)$ and $\tilde{g}_i$ can be found in~\cite{Ciuchini:2015qxb}\footnote{Due to a typographic error in eq.~2.7 of~\cite{Ciuchini:2015qxb} the denominators have $C_1$ when they should be $C_2$. We thank Christoph Bobeth and Danny Van Dyke for pointing this out.}. On the top panels of fig.~\ref{fig:comp} the blue band represents $2C_2\vert\tilde{g}_{i}^{fit}\vert$ extracted from LHCb data when constraining $\tilde{g}_{i}$ using estimates from~\cite{Khodjamirian:2010vf} below $q^2=1$ GeV$^2$ leading to the function having a strong $q^2$ dependence at larger $q^2$. This can be compared to the middle panels where the theoretical estimates from~\cite{Khodjamirian:2010vf} are not used at all and hence the functional form at higher $q^2$ is not strongly dependent on $q^2$. The bottom panels pertain to the case where we set $h_\lambda^{(2)}=0$ making $h_\lambda(q^2) = h_\lambda^{(0)} + \frac{q^2}{1\,\mathrm{GeV}^2}h_\lambda^{(1)}$. The resulting $2C_2\vert\tilde{g}_{i}^{fit}\vert$ is almost flat at larger $q^2$. This allows for it to be interpreted as a constant shift in $C_9$ which could not only be generated by QCD corrections but could also be an effect of shifts in the Wilson coefficients generated by NP at higher scales. Comparing the top and the bottom panels of fig.~\ref{fig:comp} it is evident that the functional form used to extract the unknown part necessary to fit the data is an important discriminator between possible NP and QCD contributions. We showed in our previous work that the goodness of fit to data was almost the same for both the first and the third cases. This leads us to believe that either more data is necessary to test the $q^2$ dependence of the non-factorizable contribution, a significant presence of which will point to a SM interpretation, or a complete theoretical estimation needs to be made to disentangle NP from SM. 


\section{Updated results and analysis}
\label{sec:results}

\begin{figure}[!h]
\centering
\subfigure{\includegraphics[width=.245\textwidth]{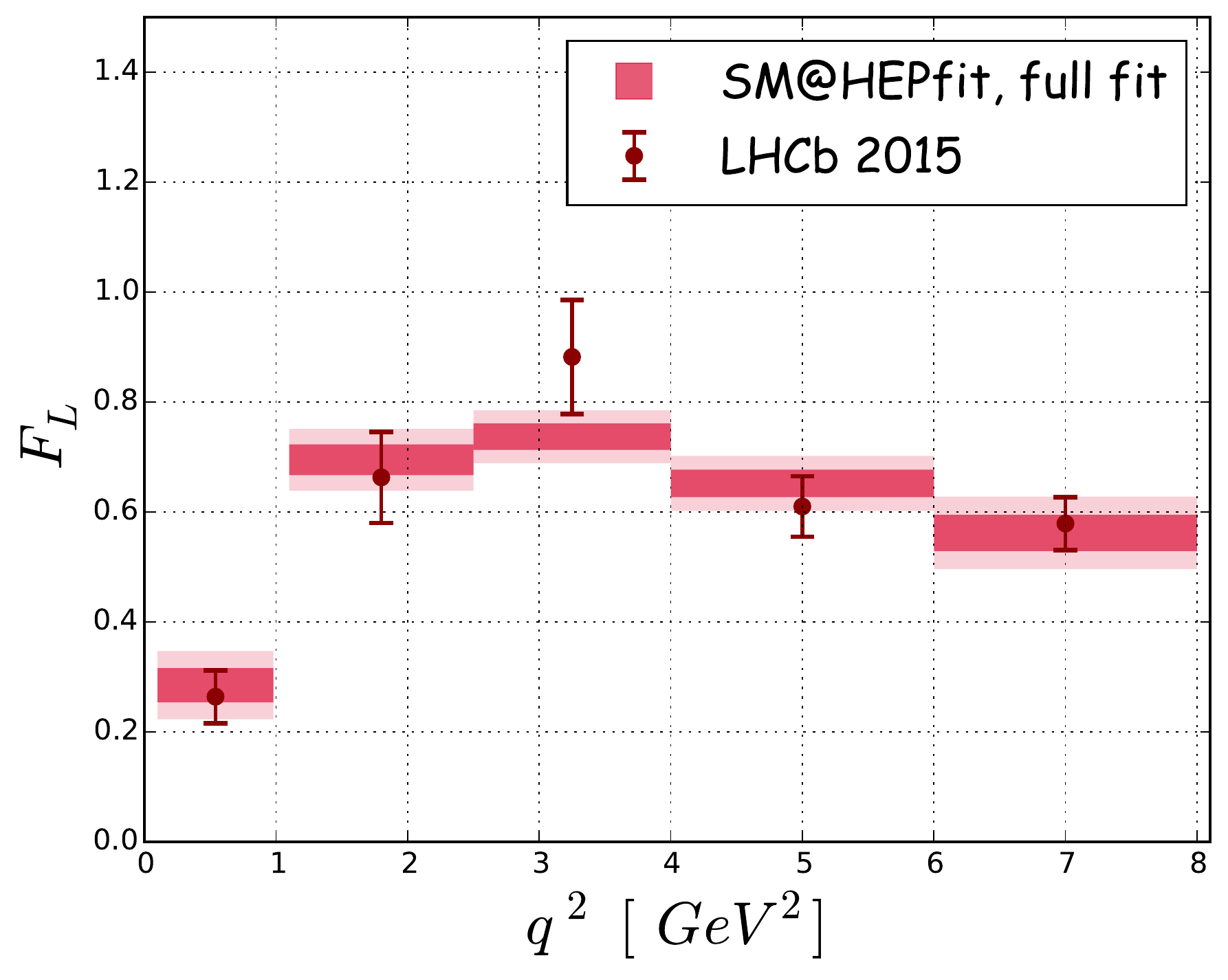}}
\subfigure{\includegraphics[width=.245\textwidth]{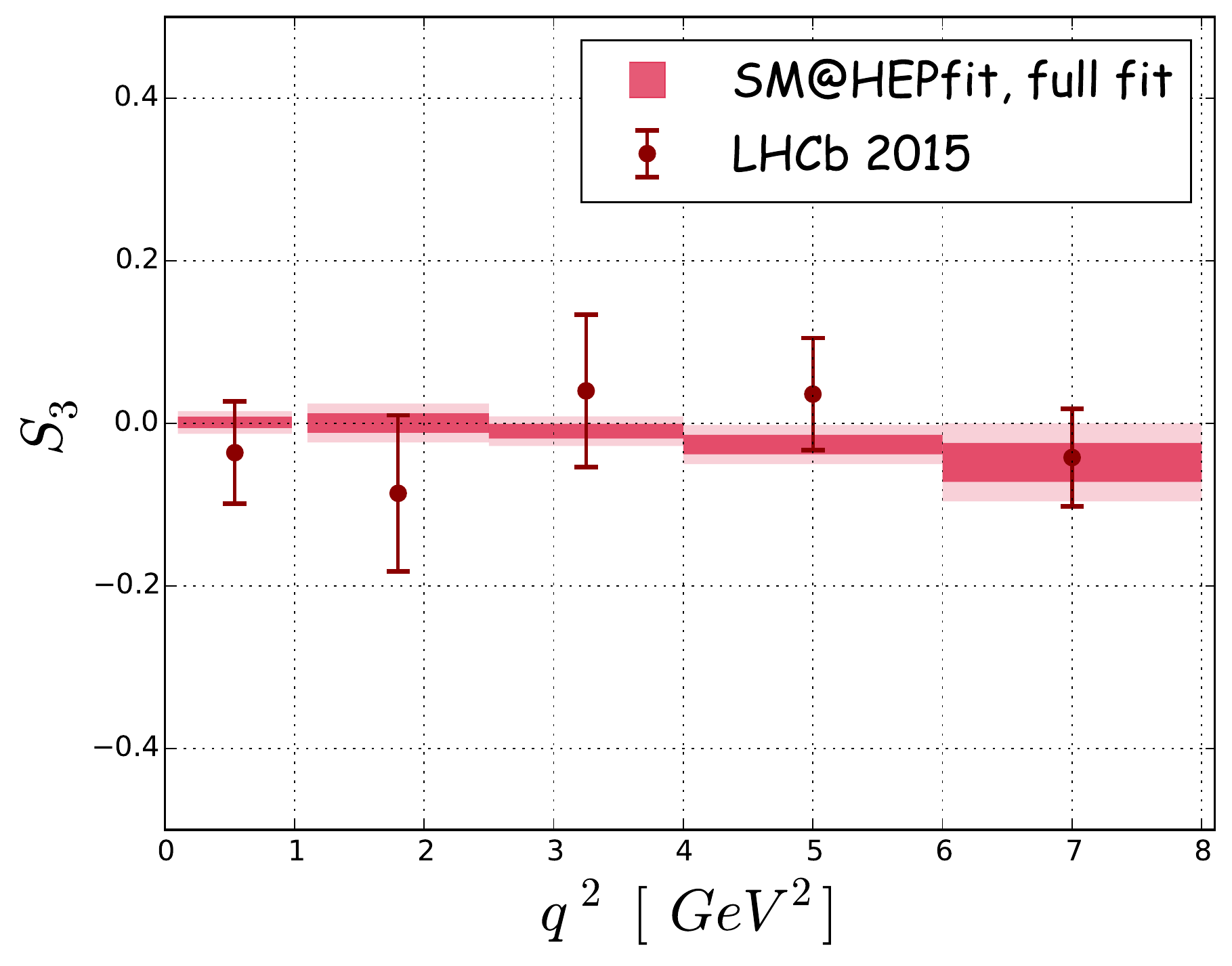}}
\subfigure{\includegraphics[width=.245\textwidth]{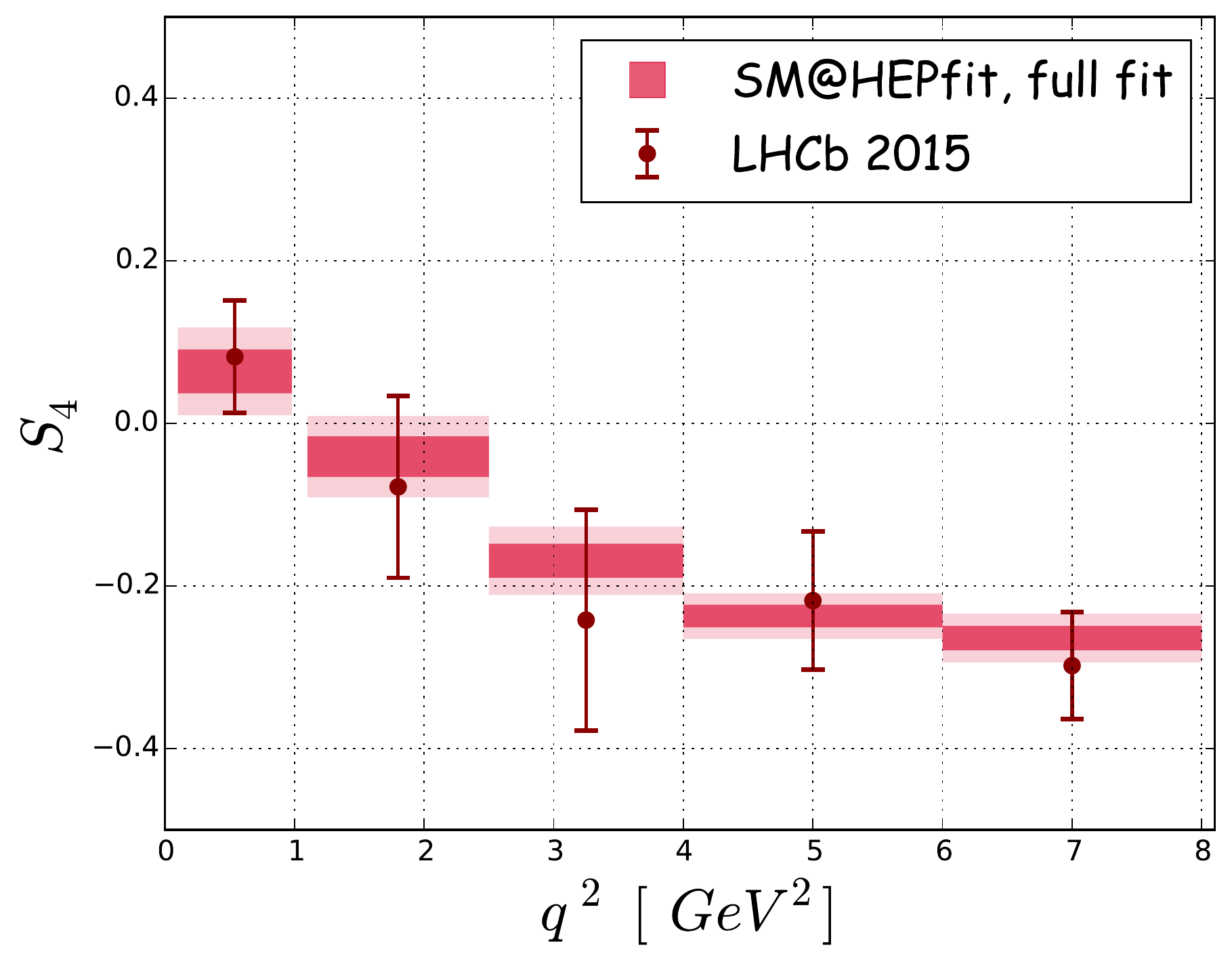}}
\subfigure{\includegraphics[width=.245\textwidth]{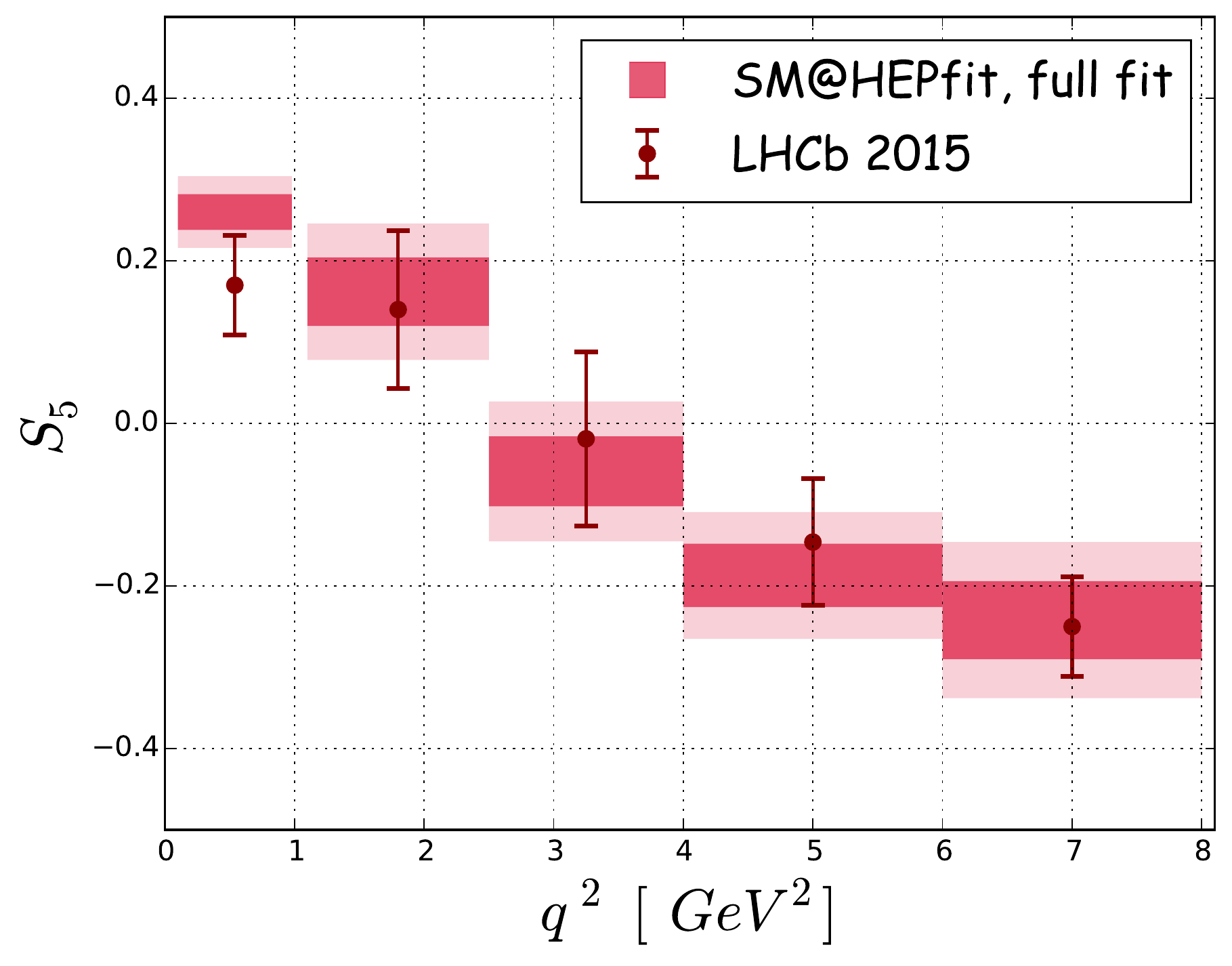}}
\subfigure{\includegraphics[width=.245\textwidth]{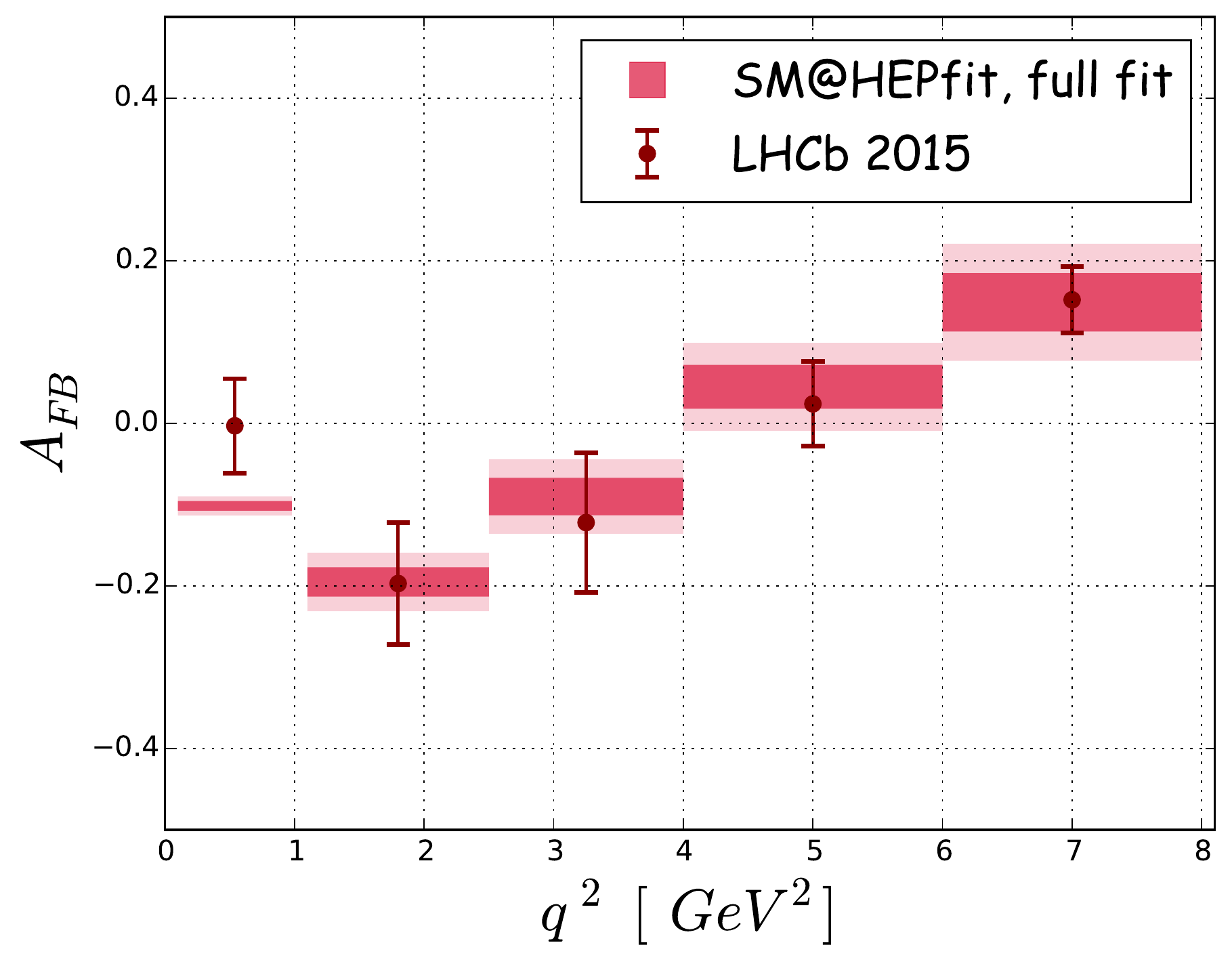}}
\subfigure{\includegraphics[width=.245\textwidth]{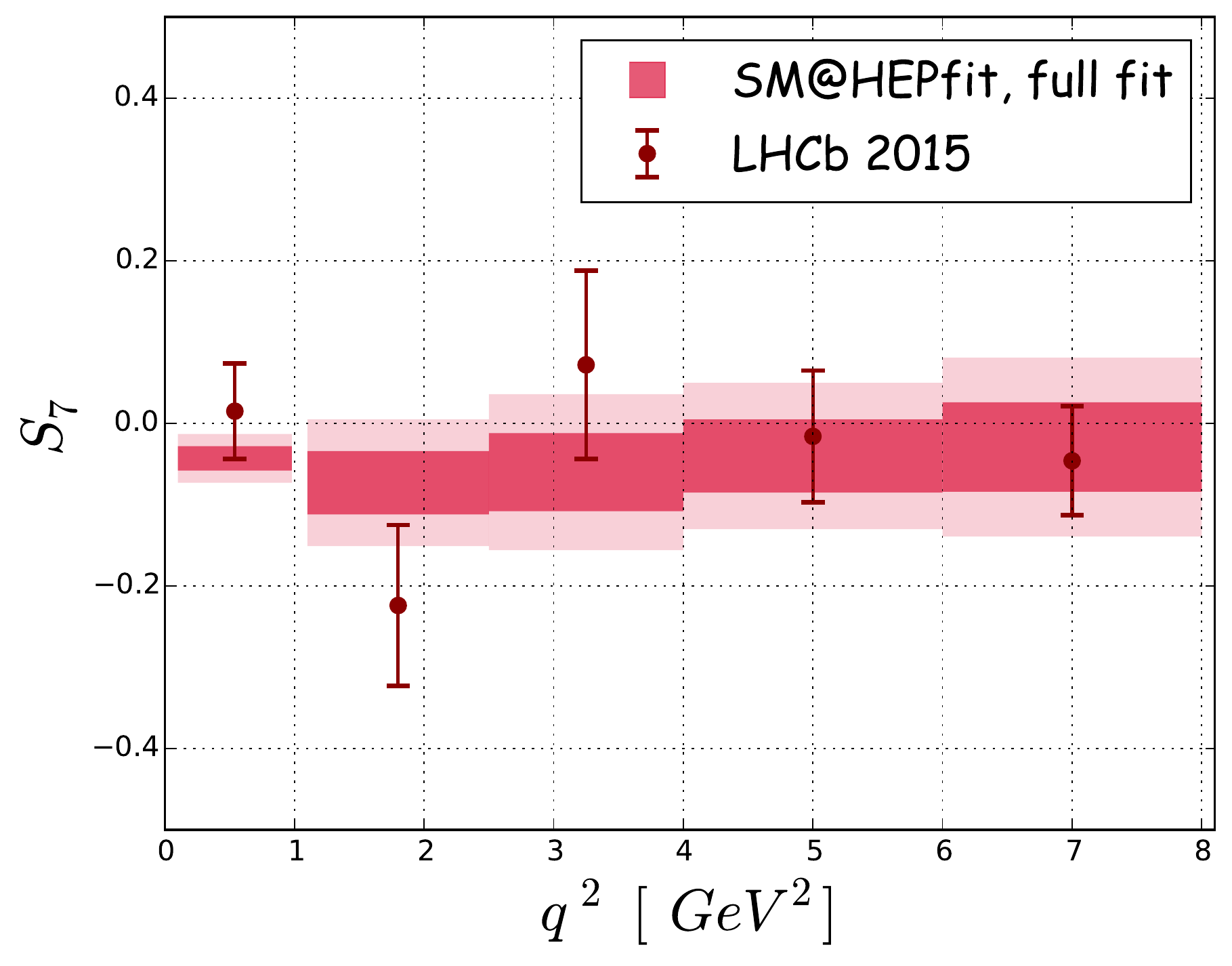}}
\subfigure{\includegraphics[width=.245\textwidth]{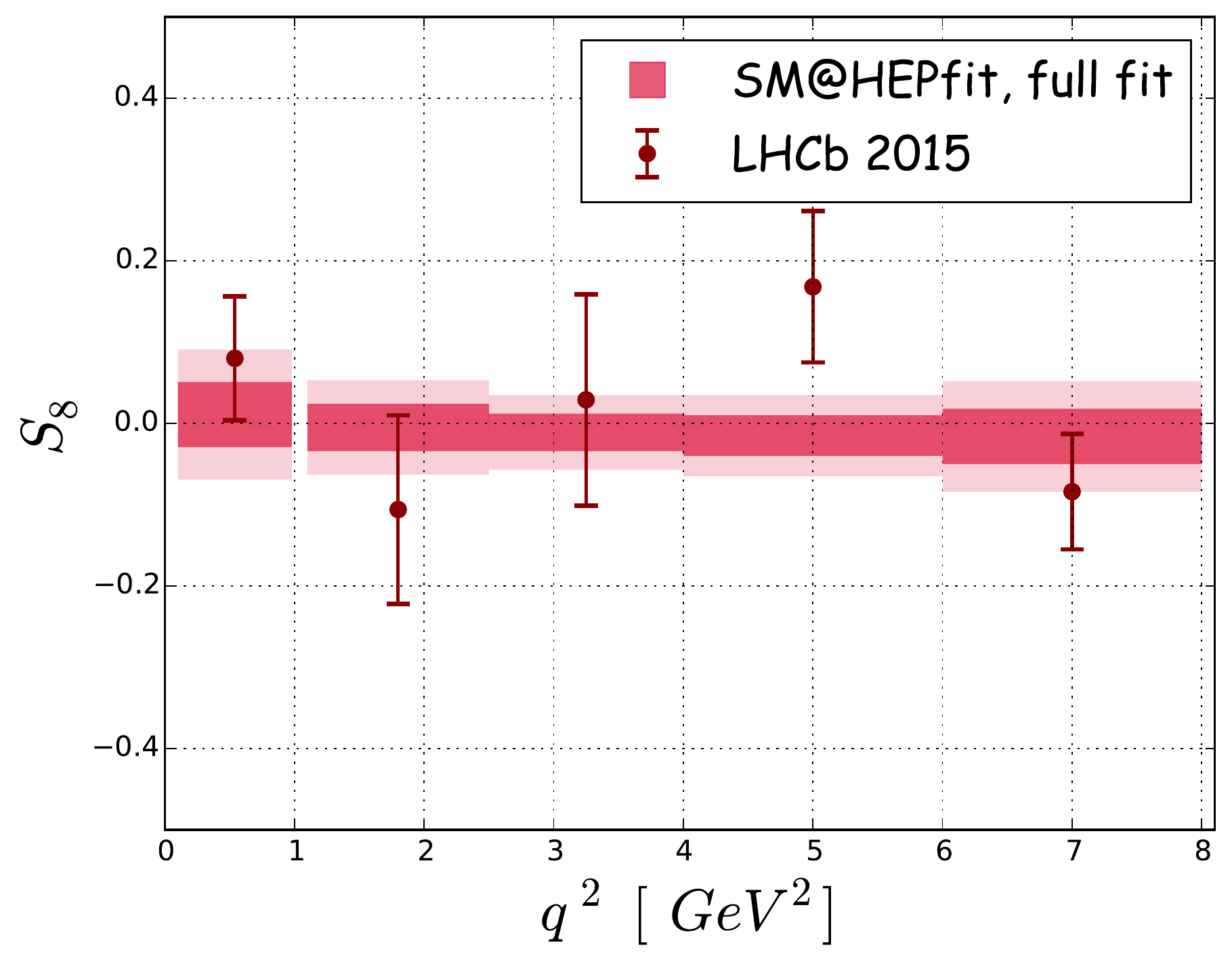}}
\subfigure{\includegraphics[width=.245\textwidth]{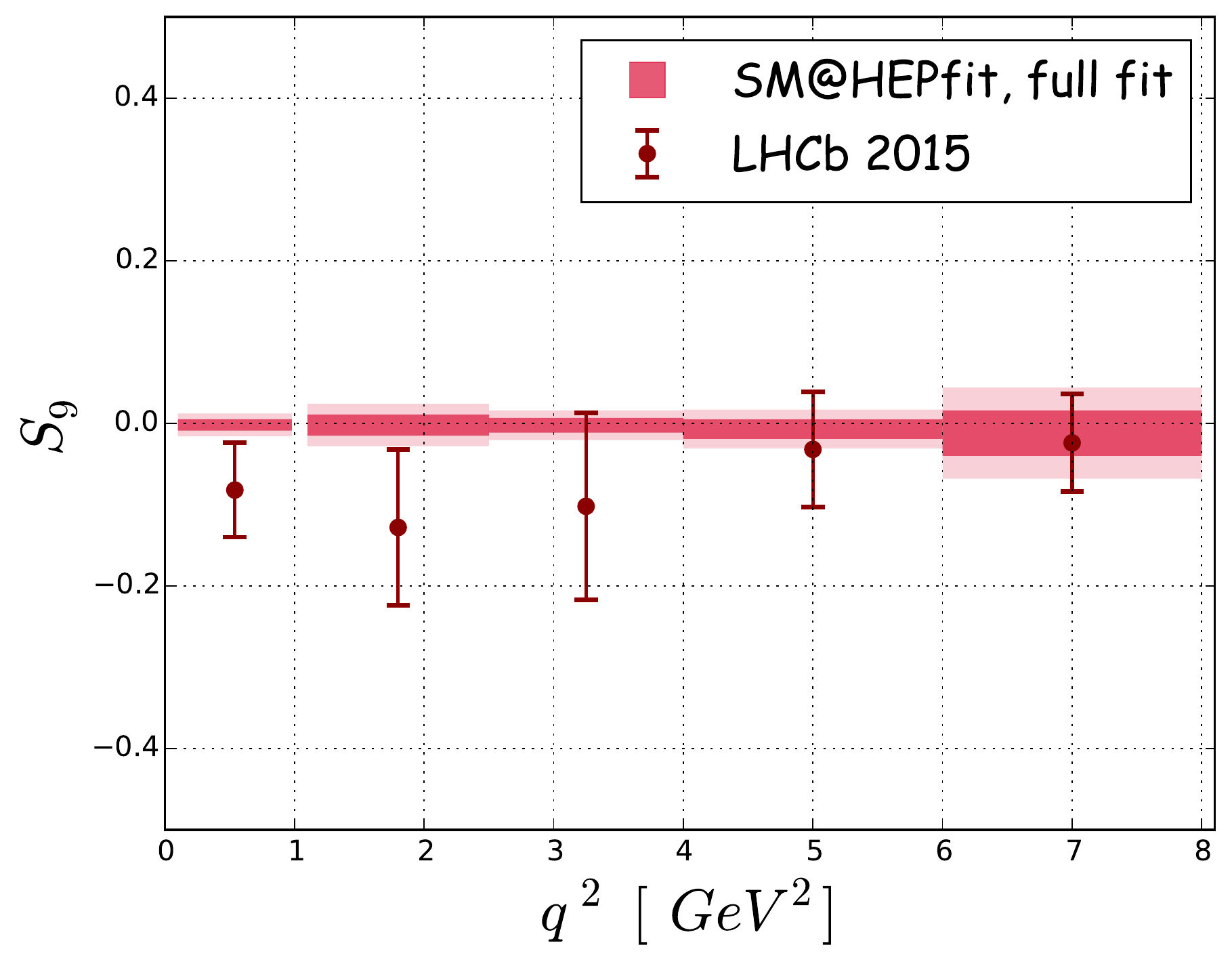}}
\caption{\textit{Results of the full fit and experimental results for
    the $B \to K^* \mu^+ \mu^-$ angular observables. Here and in the
    following, we use darker (lighter) colours for the $68\%$ ($95\%$)
    probability regions.}}
\label{fig:fullfit}
\end{figure}

Recently, the form factors calculated using LCSR were updated by the authors of~\cite{Straub:2015ica} and also combined with the form factors calculated using Lattice QCD techniques~\cite{Horgan:2015vla} which we use to update our fits. In fig.~\ref{fig:fullfit} we present the result where we use the theoretical estimates of $\tilde{g}_i$ from~\cite{Khodjamirian:2010vf} below $q^2=1$ GeV$^2$. We will not present the other scenarios presented in~\cite{Ciuchini:2015qxb} in this proceeding. Comparing tab.~\ref{tab:main} to tab.~2 in~\cite{Ciuchini:2015qxb} the fit in the first two bins are better while the results in the other bins remain mostly unchanged. Our conclusion remains the same. The hadronic contribution extracted from this fit (top panels of fig.~\ref{fig:comp}) is larger than the leading contribution in~\cite{Khodjamirian:2010vf} when one uses the dispersion relation to extrapolate to $q^2\gg 1$ GeV$^2$ (gray band). However, they are reasonable taking into account that effect close to the $c\bar{c}$ threshold can potentially be quite large. 

\begin{table}[!htbp]
\fontsize{8}{8}\selectfont
\centering
\begin{tabular}{|c|c|c|c|c|c|}
\hline
&&&&&\\[-1mm]
\textbf{$q^2$ bin [GeV$^2$]} & \textbf{Observable} & \textbf{measurement} & \textbf{full fit} & \textbf{prediction} & $\mathbf{p-value}$ \\[1mm]
\hline
&&&&&\\[-2mm]
 \multirow{9}{*}{\normalsize $ [0.1,0.98] $}
& $F_L
$ & $
\phantom{-}
0.264 \pm 0.048$ & $
\phantom{-}
0.285 \pm 0.031$ & $
\phantom{-}
0.278 \pm 0.039$ & \multirow{8}{*}{0.24}
\\
& $S_3
$ & $
-0.036 \pm 0.063$ & $
\phantom{-}
0.001 \pm 0.007$ & $
\phantom{-}
0.002 \pm 0.007$ & 
\\
& $S_4
$ & $
\phantom{-}
0.082 \pm 0.069$ & $
\phantom{-}
0.064 \pm 0.027$ & $
\phantom{-}
0.036 \pm 0.045$ & 
\\
& $S_5
$ & $
\phantom{-}
0.17 \pm 0.061$ & $
\phantom{-}
0.260 \pm 0.022$ & $
\phantom{-}
0.277 \pm 0.027$ & 
\\
& $ A_{FB}
$ & $
-0.003 \pm 0.058$ & $
-0.101 \pm 0.006$ & $
-0.103 \pm 0.007$ & 
\\
& $S_7
$ & $
\phantom{-}
0.015 \pm 0.059$ & $
-0.043 \pm 0.015$ & $
-0.040 \pm 0.017$ & 
\\
& $S_8
$ & $
\phantom{-}
0.08 \pm 0.076$ & $
\phantom{-}
0.011 \pm 0.040$ & $
-0.004 \pm 0.046$ & 
\\
& $S_9
$ & $
-0.082 \pm 0.058$ & $
-0.002 \pm 0.007$ & $
-0.002 \pm 0.007$ & 
\\
\cline{2-6}
&&&&&\\[-2mm]
& $ P_5'
$ & $
\phantom{-}
0.387 \pm 0.142$ & $
\phantom{-}
0.728 \pm 0.072$ & $
\phantom{-}
0.785 \pm 0.095$ & 0.020
\\[.5mm]
\hline
&&&&&\\[-2mm]
 \multirow{9}{*}{\normalsize $ [1.1,2.5] $}
& $F_L
$ & $
\phantom{-}
0.663 \pm 0.083$ & $
\phantom{-}
0.695 \pm 0.028$ & $
\phantom{-}
0.694 \pm 0.032$ & \multirow{8}{*}{0.64}
\\
& $S_3
$ & $
-0.086 \pm 0.096$ & $
\phantom{-}
0.001 \pm 0.012$ & $
\phantom{-}
0.002 \pm 0.012$ & 
\\
& $S_4
$ & $
-0.078 \pm 0.112$ & $
-0.041 \pm 0.025$ & $
-0.048 \pm 0.030$ & 
\\
& $S_5
$ & $
\phantom{-}
0.14 \pm 0.097$ & $
\phantom{-}
0.162 \pm 0.042$ & $
\phantom{-}
0.178 \pm 0.052$ & 
\\
& $ A_{FB}
$ & $
-0.197 \pm 0.075$ & $
-0.195 \pm 0.018$ & $
-0.197 \pm 0.021$ & 
\\
& $S_7
$ & $
-0.224 \pm 0.099$ & $
-0.073 \pm 0.039$ & $
-0.050 \pm 0.045$ & 
\\
& $S_8
$ & $
-0.106 \pm 0.116$ & $
-0.005 \pm 0.029$ & $
-0.005 \pm 0.032$ & 
\\
& $S_9
$ & $
-0.128 \pm 0.096$ & $
-0.002 \pm 0.013$ & $
\phantom{-}
0.002 \pm 0.013$ & 
\\
\cline{2-6}
&&&&&\\[-2mm]
& $ P_5'
$ & $
\phantom{-}
0.298 \pm 0.212$ & $
\phantom{-}
0.363 \pm 0.090$ & $
\phantom{-}
0.400 \pm 0.110$ & 0.67
\\[.5mm]
\hline
&&&&&\\[-2mm]
 \multirow{9}{*}{\normalsize $ [2.5,4] $}
& $F_L
$ & $
\phantom{-}
0.882 \pm 0.104$ & $
\phantom{-}
0.737 \pm 0.024$ & $
\phantom{-}
0.728 \pm 0.027$ & \multirow{8}{*}{0.80}
\\
& $S_3
$ & $
\phantom{-}
0.04 \pm 0.094$ & $
-0.010 \pm 0.009$ & $
-0.011 \pm 0.010$ & 
\\
& $S_4
$ & $
-0.242 \pm 0.136$ & $
-0.169 \pm 0.021$ & $
-0.169 \pm 0.022$ & 
\\
& $S_5
$ & $
-0.019 \pm 0.107$ & $
-0.059 \pm 0.043$ & $
-0.056 \pm 0.049$ & 
\\
& $ A_{FB}
$ & $
-0.122 \pm 0.086$ & $
-0.090 \pm 0.023$ & $
-0.093 \pm 0.025$ & 
\\
& $S_7
$ & $
\phantom{-}
0.072 \pm 0.116$ & $
-0.060 \pm 0.048$ & $
-0.080 \pm 0.054$ & 
\\
& $S_8
$ & $
\phantom{-}
0.029 \pm 0.13$ & $
-0.011 \pm 0.023$ & $
-0.010 \pm 0.024$ & 
\\
& $S_9
$ & $
-0.102 \pm 0.115$ & $
-0.002 \pm 0.009$ & $
-0.003 \pm 0.010$ & 
\\
\cline{2-6}
&&&&&\\[-2mm]
& $ P_5'
$ & $
-0.077 \pm 0.354$ & $
-0.140 \pm 0.100$ & $
-0.130 \pm 0.110$ & 0.89
\\[.5mm]
\hline
&&&&&\\[-2mm]
 \multirow{9}{*}{\normalsize $ [4,6] $}
& $F_L
$ & $
\phantom{-}
0.61 \pm 0.055$ & $
\phantom{-}
0.652 \pm 0.025$ & $
\phantom{-}
0.661 \pm 0.029$ & \multirow{8}{*}{0.52}
\\
& $S_3
$ & $
\phantom{-}
0.036 \pm 0.069$ & $
-0.026 \pm 0.012$ & $
-0.026 \pm 0.013$ & 
\\
& $S_4
$ & $
-0.218 \pm 0.085$ & $
-0.237 \pm 0.014$ & $
-0.234 \pm 0.017$ & 
\\
& $S_5
$ & $
-0.146 \pm 0.078$ & $
-0.187 \pm 0.039$ & $
-0.209 \pm 0.045$ & 
\\
& $ A_{FB}
$ & $
\phantom{-}
0.024 \pm 0.052$ & $
\phantom{-}
0.045 \pm 0.027$ & $
\phantom{-}
0.061 \pm 0.032$ & 
\\
& $S_7
$ & $
-0.016 \pm 0.081$ & $
-0.040 \pm 0.045$ & $
-0.046 \pm 0.054$ & 
\\
& $S_8
$ & $
\phantom{-}
0.168 \pm 0.093$ & $
-0.015 \pm 0.025$ & $
-0.026 \pm 0.026$ & 
\\
& $S_9
$ & $
-0.032 \pm 0.071$ & $
-0.007 \pm 0.012$ & $
-0.011 \pm 0.013$ & 
\\
\cline{2-6}
&&&&&\\[-2mm]
& $ P_5'
$ & $
-0.301 \pm 0.160$ & $
-0.396 \pm 0.085$ & $
-0.446 \pm 0.099$ & 0.44
\\[.5mm]
\hline
&&&&&\\[-2mm]
 \multirow{9}{*}{\normalsize $ [6,8] $}
& $F_L
$ & $
\phantom{-}
0.579 \pm 0.048$ & $
\phantom{-}
0.562 \pm 0.033$ & $
\phantom{-}
0.511 \pm 0.069$ & \multirow{8}{*}{0.82}
\\
& $S_3
$ & $
-0.042 \pm 0.06$ & $
-0.048 \pm 0.024$ & $
-0.006 \pm 0.053$ & 
\\
& $S_4
$ & $
-0.298 \pm 0.066$ & $
-0.264 \pm 0.015$ & $
-0.225 \pm 0.036$ & 
\\
& $S_5
$ & $
-0.25 \pm 0.061$ & $
-0.242 \pm 0.048$ & $
-0.166 \pm 0.098$ & 
\\
& $ A_{FB}
$ & $
\phantom{-}
0.152 \pm 0.041$ & $
\phantom{-}
0.149 \pm 0.036$ & $
\phantom{-}
0.111 \pm 0.076$ & 
\\
& $S_7
$ & $
-0.046 \pm 0.067$ & $
-0.029 \pm 0.055$ & $
\phantom{-}
0.020 \pm 0.110$ & 
\\
& $S_8
$ & $
-0.084 \pm 0.071$ & $
-0.016 \pm 0.034$ & $
\phantom{-}
0.037 \pm 0.053$ & 
\\
& $S_9
$ & $
-0.024 \pm 0.06$ & $
-0.012 \pm 0.028$ & $
\phantom{-}
0.018 \pm 0.047$ & 
\\
\cline{2-6}
&&&&&\\[-2mm]
& $ P_5'
$ & $
-0.505 \pm 0.124$ & $
-0.491 \pm 0.098$ & $
-0.340 \pm 0.200$ & 0.48
\\[.5mm]
\hline
\hline&&&&&\\[-2mm]
 $ [0.1,2] $ & 
\multirow{3}{*}{\normalsize $ {\rm BR } \cdot 10^7 $}
 & $ 0.58 \pm 0.09$ & $
0.65 \pm 0.04$ & $
0.66 \pm 0.04$ & 0.42
\\
 $ [2,4.3] $ & 
 & $ 0.29 \pm 0.05$ & $
0.33 \pm 0.03$ & $
0.35 \pm 0.04$ & 0.35
\\
 $ [4.3,8.68] $ & 
 & $ 0.47 \pm 0.07$ & $
0.45 \pm 0.05$ & $
0.46 \pm 0.10$ & 0.93
\\
\hline

&&&&&\\[-2mm]
& { \normalsize $ {\rm BR }_{B \to K^* \gamma} \cdot 10^5 $}
& $ 4.33 \pm 0.15$ & $
4.35 \pm 0.14$ & $
4.73 \pm 0.54$ & 0.48
\\[1mm]
\hline
\end{tabular}
\caption{\footnotesize\textit{Experimental results (with symmetrized errors), results from the full fit, predictions and $p$-values for $B \to K^* \mu^+ \mu^-$ BR's and angular observables. The predictions are obtained removing the corresponding observable from the fit. For the angular observables, since their measurements are correlated in each bin, we remove from the fit the experimental information on all angular observables in one bin at a time to obtain the predictions. See~\cite{Ciuchini:2015qxb} for details. We also report the results for ${\rm BR}(B \to K^*\gamma)$ and for the optimized observable $P_5^\prime$. The latter is however not explicitly used in the fit as a constraint, since it is not independent of $F_L$ and $S_5$.}}
\label{tab:main}
\end{table}

\section{Conclusion}
\label{sec:conclusion}

In this proceeding, we critically examine the interpretation of the theoretical uncertainty coming from nonfactorizable corrections in the region of $q^2 \lesssim 4 m_c^2$ for the channel $B \to K^* \ell^+ \ell^-$. Using all available theoretical estimates we extract both the size and the shape of these contributions and find no significant discrepancy with SM expectations. This requires the presence of sizable, yet perfectly acceptable, nonfactorizable power corrections. Assuming the validity of the QCD sum rules estimate of these power corrections at $q^2 \leq 1$ GeV$^2$, we observe a $q^2$ dependence of the nonfactorizable contributions which disfavours their interpretation as a shift of the SM Wilson coefficients. A fit performed without using any theoretical estimate of the nonfactorizable corrections yields a range for these contributions larger than, but in the same ballpark of, the QCD sum rule calculation. In this case, unfortunately, no conclusion on the presence of $q^4$ terms in $h_\lambda$ can be drawn. We also present the case where we set $h_\lambda^{(2)}=0$, hence removing the $q^4$ dependence in the nonfactorizable corrections which results in a flat curve over almost the entire range of $q^2$ making it tantamount to an interpretation in terms of SD contribution from higher scales. We conclude that no evidence of CP-conserving NP contributions to the Wilson coefficients $C_{7,9}$ can be inferred unless a theoretical breakthrough allows us to obtain an accurate estimate of nonfactorizable power corrections and to disentangle possible NP contributions from hadronic uncertainties. Nevertheless, an improved set of measurements could possibly clarify the issue of the $q^2$ dependence of the nonfactorizable contributions along with detailed analyses of the radiative modes $B \to V\gamma$~\cite{Paul:2016urs}.

\bibliographystyle{JHEP}
\bibliography{ICHEP2016_btos}
%

\end{document}